\documentstyle[12pt,aasms4]{article}

\slugcomment{Submitted to PASP}

\begin{document}

\title{ISO Observations of the 53W002 Group at 6.7$\mu$m: 
In Search of the Oldest Stellar Populations at $z=2.4$
\footnote{Based on observations with ISO, an ESA project with instruments 
funded by ESA Member States (especially the PI countries: France, Germany, 
the Netherlands and the United Kingdom) and with the participation of ISAS 
and NASA.}}

\author{William C. Keel\altaffilmark{2} and Wentao Wu\altaffilmark{3}}
\affil{Department of Physics and Astronomy, University of Alabama,
Box 8709324, Tuscaloosa, AL 35487-0324; keel@bildad.astr.ua.edu}

\author{Paul P. van der Werf}
\affil{Sterrewacht Leiden, Postbus 9513, 2300 RA Leiden, The Netherlands;
pvdwerf@strw.leidenuniv.nl}

\author{Rogier A. Windhorst}
\affil{Dept. of Physics and Astronomy, Arizona State University,
Box 1504, Tempe, AZ 85287; Rogier.Windhorst@asu.edu}

\author{James S. Dunlop}
\affil{Royal Observatory Edinburgh, Blackford Hill, Edinburgh EH9 3HJ, UK;
J.Dunlop@roe.ac.uk}

\author{Stephen A. Eales}
\affil{Department of Physics and Astronomy, University of Wales, Cardiff,
5, The Parade, Cardiff CF24 3YB, UK; Steve.Eales@astro.cf.ac.uk}

\author{Ian Waddington}
\affil{Astronomy Centre, Dept. of Physics \& Astronomy,
University of Sussex, Falmer,
Brighton BN1 9QH, U.K.; I.Waddington@sussex.ac.uk}

\author{Martha Holmes\altaffilmark{4}}
\affil{Dept. of Physics and Astronomy, Box 1803-B, Vanderbilt
University, Nashville, TN 37235; martha.j.holmes@vanderbilt.edu}

\altaffiltext{2} {Visiting Astronomer at the Infrared Telescope Facility,
 which is operated
by the University of Hawaii under Cooperative Agreement no. NCC 5-538
with the National Aeronautics and Space Administration, Office of Space
Science, Planetary Astronomy Program. }
 
\altaffiltext{3} {Present address: Department of Astronomy \& Astrophysics,
525 Davey Lab, Pennsylvania State University, University Park, PA 16802}

\altaffiltext{4} {NSF Research Experiences for Undergraduates participant at the
University of Alabama}

\begin{abstract}
We present a deep ISO observation at 6.7$\mu$m
of the 53W002 group of galaxies and
AGN at $z=2.4$. This approximately samples the emitted
$K$ band. The faint, blue star-forming objects are not detected,
as expected from their very blue color across the emitted optical and
UV. However, 53W002 itself is detected at the $\approx 3 \sigma$ level, with
an emitted $V-K$ color appropriate for a population formed starting
at $z=3.6$--7.0 with most likely value $z=4.7$. 
This fits with shorter-wavelength data suggesting
that the more massive members of this group, which may
all host AGN, began star formation earlier in deeper potential
wells than the compact Lyman $\alpha$ emission objects. Two
foreground galaxies are detected, as well as several stars.
One additional 6.7$\mu$m source closely coincides with an
optically faint galaxy, potentially at $z=2-3$.
The overall source counts are consistent (within the errors of
such a small sample) with the other ISO deep fields at $6.7 \mu$m.
 
\end{abstract}


\section{Introduction}

As our understanding of galaxies at high redshifts improves,
it becomes important to be able to compare them
in increasing detail with nearby systems, as well as with evolutionary models.
This has driven a number of studies of these objects in the
infrared to sample the emitted optical range with its
many diagnostic spectral features for accessible redshifts.
Near-infrared spectroscopy has revealed the locally well-studied
[O III], [N II], and Balmer emission lines, and in a few cases
spectral breaks that indicate an age for the dominant stellar 
population (Eales \& Rawlings 1993, Pettini et al. 1998, 2001,
Teplitz et al. 2000, Motohara et al. 2001). Multiband photometry 
has provided important
information as well, especially given the difficulty of obtaining
spectra of such faint objects against the high background of
ground-based observations.

Long wavelength baselines improve estimates of the spectral shape, and
consequently effective ages, of stellar populations, since the sensitivity
to a minority population of young stars and to dust effects are both reduced.
We present here deep ISO observations of a group of objects at
$z=2.4$, corresponding to 2$\mu$m in the emitted frame, and thus offering
a sensitive probe of the oldest stellar populations
in these objects. At high redshifts, even crude limits on age
of a stellar population become very sensitive measures of formation
time, and can sometimes limit cosmological parameters, as the
redshift/age relation becomes stronger (Stockton et al. 1995,
Dunlop et al. 1996, Spinrad et al. 1997). 

We targeted the region around the radio galaxy 53W002 at $z=2.39$
(Windhorst et al. 1991) for ISO observations in the emitted near-infrared. 
This system has been found to be part of
a grouping or cluster extending over at least 7' (3.5 Mpc for
the WMAP cosmology with
${\rm H}_0=71$ km s$^{-1}$ Mpc$^{-1}$, $\Omega_M = 0.27$, and
$\Lambda$-dominated flat geometry),
including 5 additional AGN and 8 star-forming objects detected
through narrow Lyman $\alpha$ emission (Pascarelle et al. 
1996, Keel et al. 1999). 
These fainter Lyman $\alpha$ sources are especially interesting as
candidate protogalactic objects (of rather low mass).
They are quite small, with effective radii in the
emitted ultraviolet of 0\farcs1--0\farcs2 (1--2 kpc), and some are so closely
paired as to suggest a short timescale against merging, as expected from
hierarchical scenarios. This field offered the chance for uniquely
deep detections or limits on such objects, since we could observe
7 such objects at once and derive an average flux or limit several times
deeper than possible for an individual target. While it was clear
that detections of young blue populations would be at the very limit of 
ISOCAM's sensitivity, the opportunity to measure so many high-redshift targets 
(both AGN and star-forming
systems) at once was quite attractive.

More recent NICMOS observations (Keel et al. 2002) show that the
star-forming objects are very blue (and presumably thus quite
poor in both dust and metals), so that their detections at
6.7$\mu$m would be unlikely. Still, these data are valuable
in setting an explicit limit on the emitted $R-K$ colors,
and in the likely detection of 53W002 itself.
This is also one of the handful of deepest ISO images at this
wavelength, in a field for which supporting data from the optical, near-IR,
and radio regimes are available, so the source counts and identifications
are of interest in themselves. For low-redshift galaxies,
this wavelength falls between photospheric starlight and the hottest dust
emission, making the detected population relatively sensitive to
high-redshift galaxies.

\section{Observations}

\subsection{ISO 6.7$\mu$m Data}

The 53W002 field was observed with 
ISOCAM (Cesarsky et al. 1996) on
orbits 500 and 521. The LW2 filter, with half-power transmission
points at 5.0 and 8.5 $\mu$m and effective wavelength 6.7 $\mu$m for
a flat spectrum, was used to approximate the emitted $K$ band at
$z=2.4$ (the effective emitted-frame wavelength actually corresponds to
1.97$\mu$m). Since we were interested in 11 objects in the
53W002 group (including three foreground and background Lyman
$\alpha$ emitters at $z=2.05-2.7$), a tradeoff had to be made between
covering a wide enough field and high point-source sensitivity. We used the 
3" pixel scale,
undersampling the LW2 PSF in the interest of covering a larger
region at once. The final spatial resolution will be somewhat worse
than this, since data from many offset pointings are stacked.
Two levels of pointing offsets were used for these data. At each of 21
positions, $3 \times 3$ sets of $10 \times 10.08$-second exposures
were done, with the $3 \times 3$ pattern using offsets of 5 pixels
(15"). A set of 21 of these pointings was laid out to cover the
area spanning most of the known objects around 53W002, with 45"
spacing in both celestial coordinates. To provide scheduling
flexibility while assuring that the right regions were observed,
these locations were specified in celestial rather than spacecraft
coordinates, with the result that the detector orientation is
skewed to the pointing grid (by about $109^\circ$ for orbit 500
and $129^\circ$ for orbit 521, with rotations of $0.25^\circ$ or
less within each orbit). The overlap of these observations is shown in
Fig. 1. To minimize the time used in microslews
between these positions, they were observed in a
boustrophedonic order. For clarity in describing the reductions, 
we will refer to each $3 \times 3$ set of exposures as a raster.

Before each raster observation, a brief exposure was made using the 
LW1 filter, taking advantage of experience which suggested that the
LW detector response to transients could be improved by a change in
the background level. The total exposure time in LW2 was 19051 seconds
(5.3 hours), of which a typical useful total at a given sky position
is 13000 seconds in the inner part of the field. The region of best
exposure is roughly circular with a radius of 90", centered at (J2000)
$\alpha$ = 17:14:13.0, $\delta$ = +50:15:57. There is considerable 
structure at the 10\% level in the 
exposure pattern even within this region, but this amplitude
does not strongly change the detection limit. 
In comparison to other ISO surveys at this wavelength, this study is thus
substantially deeper than the CFRS field examined by
Flores et al. (1999), 
comparable in exposure to the LW2 data on the Hubble Deep Field
presented by Serjeant et al. (1997) and about 3 times shorter than the very 
deep SSA13 and Lockman Hole observations by Taniguchi et al. (2002).


Reduction of these data began using the ISOCAM CIA package (v4.0,
April 2000 release, described in Delaney 2000). Its routines were used to 
clean the data stream
of incomplete and otherwise invalid exposures, correct for the dark
pattern by standard library frames, deglitch, and stabilize the
individual exposures. Deglitching, which corrects for 
long-lasting radiation-induced
events, was found to be most effective with
the multiresolution median algorithm in CIA. The stabilization step corrects
for the substantial time constant of the detector in responding to a
change in illumination (such as happens at each change of pointing).
The data for each raster were combined to form a single mosaic,
rejecting pixels masked as uncorrectable in the earlier stage,
and each mosaic was projected onto a common celestial-coordinate
frame, expanded to 1\farcs0 pixels using the C routine {\it project} distributed
with the CIA package. This expansion preserved the intensity-flux
calibration in mJy per solid angle.
The number of glitches not rejected by these techniques was still
substantial, so we carried out further processing and masking using
the {\tt combine} and {\tt imedit} tasks
within IRAF. 
 
Residual flat-field structures were apparent in the raster products,
so these were corrected using a median frame for each orbit's data.
Drifts in background level between rasters were also removed. 
Glitches that remained after the CIA processing were identified by
comparison of spatially overlapping rasters, and masked before
combining. The best results were
found by masking the outer parts of each raster as well, leaving those
regions which were included in at least 6 of the 9 constituent
observations (a region 111" square in each case). The final
data mosaic combined all these masked raster products with
$3 \sigma$ rejection. The effective resolution of this mosaic 
should be slightly better than 6" FWHM, a combination of the input
3" pixel scale and stacking of variously offset observations. This
value varies significantly depending on a source's exact location
with respect to the observed pixel grid; some star images have a FWHM
as small as 4".

The mosaicked image still shows residual variations in blank-sky
level, associated with slight slope changes from one observation
to the next. We removed these using a running $11 \times 11$-arcsecond
median filter, which substantially improved the image uniformity
and reliability of aperture fluxes while not altering image structure
on relevant scales for our small targets. The detection threshold was
assessed through statistics of individual pixel values, and by the statistics of
detected positive and negative peaks after smoothing by a 
Gaussian effective PSF with 5-arcsecond FWHM. Nine ISO sources
are robust to the details of selection aperture and measuring radius
within the 4--8" range, and to whether single discrepant pixels
are rejected or not. The effective 3$\sigma$ flux limit across
the central high-exposure area is about 0.027 mJy, below which the
non-Gaussian noise distribution prevents us from saying anything
useful about the number of sources. We also estimated the mean noise
by photometry with the same 5" radius on a set of (blank) positions
derived from the object locations by 10--20" offsets in each coordinate.
The mean $1 \sigma$ error found in this way is 0.010 mJy, in good agreement 
with the estimate from statistics of positive and negative fluctuations.
 
The nominal astrometric positioning of ISOCAM frames is known to
change with motions of the lens wheel (Blommaert et al. 2001), typically
by 2-3 pixels. The only independent way for us to check the coordinate
system is through the bright interacting galaxy 53W003 at $z \simeq 0.05$, 
near the north
edge of the mosaic. The best match between its optical and mid-IR structures
comes for a 6" shift in each coordinate, such that the actual coordinates
of an object
fall SE of the values given by the nominal ISOCAM pointing. 
This offset also aligns two
much fainter sources with relatively bright
red stars, and two known galaxies with their optical
locations, so we adopt this astrometric location. The mean
offsets, in the sense of corrections to be applied to the
nominal ISO header coordinates, are $\Delta \alpha = +0.62^s$,
$\Delta \delta = -5.4''$, which have been applied to
the listed source positions.
The possibility remains of a systematic shift between the two orbits'
data, which we can limit by noting that substacks of the region of
53W002 itself, where the two data sets overlap, both recover consistent
positions of ISO sources (within the limits set by their flux levels). 
Any differential offset
between the two sets is too small to be detected with such a faint target. 
This also limits any possible offsets associated with lens-wheel
rotations which accompanied some of the interspersed LW1 ``reset"
exposures.

We carried out tests for the reality of possible sources in this
field at various stages in the processing, because of the large
number of transient events and the availability of many redundant
measurements at most locations of interest. Possible bright sources
were examined in each raster including their locations, and
any interfering glitches were masked before further combination 
of the rasters. The sources remaining in the final mosaic image
(Fig. 2) pass these straightforward tests. For 53W002 itself, we examined
various subsets of the data, none of which shows any problem that 
could in itself mimic a source detection.

For evaluating source counts, the most conservative interpretation
is that the LW2 sources with plausible optical, radio, or submillimeter 
counterparts are
real, and an upper bound is that all the statistically significant
($> 3 \sigma$) sources are real. Given the behavior of
the detectors and radiation environment, it is hard to pin
them down more tightly, so we will consider both extremes in 
counting the overall source population.

The list of detected sources is given in Table 1. Only three
bright sources are listed from outside the 90" radius of full exposure;
53W003, a red star, and one additional possible star (object 2), listed at the top
of the table. The remainder are from the central area
of deep exposure, and form a 3$\sigma$ cut, as determined by taking the 
strongest peaks after convolution with a Gaussian function of
5" FWHM. One source is a
good match for the position of a moderate-redshift spiral galaxy,
within the WFPC2 frames of Pascarelle et al. (1996).

The ISO detectors' gain varied with such factors as
the history of energetic particle flux. Accordingly, we
have checked the flux calibration for consistency with stellar
atmosphere models, as was done by the HDF-South survey as well
(Oliver et al. 2002). In this case, we used catalogs of stellar magnitudes
from $B$ through $K$, taken from images at Palomar
(Neuschaefer \& Windhorst 1995), Kitt Peak
(Keel et al. 1999), and the NASA Infrared Telescope Facility (section 2.2).
These were supplemented by B and R magnitudes from the USNO A-2
catalog of the Palomar Sky Survey (Monet 1998, Monet et al. 2001)
for the brightest stars, which
were saturated in the deep CCD images.
The optical and near-IR fluxes were used to predict a flux
at 6.7$\mu$m for comparison with the derived ISO values. Nondetections
are important as well as detections, verifying the detection threshold.
Rare stars with dust shells will appear too bright in the mid-IR compared
to the photospheric prediction, but for a group of stars this
should not change the
derived flux limit. As it happened, the red star to the NE of 53W002
(number 6 in Table 1 and Fig. 3) was especially important, having a color-derived
spectral type of M2 V. The stellar data are consistent with the nominal
ISO LW2 flux scale, so we had no need for any correction and adopt source
fluxes ``as is". This comparison limits any systematic variation in the
ISO flux scale to the $\pm 30$\% level.

\subsection{Supporting Data}

We have made use of previously reported data from the ground
and space, including WFPC2
images from Pascarelle et al. (1996, 1998), NICMOS images in F110W and F160W
filters at
1.1 and 1.6$\mu$m (Keel et al. 2002), wide-field optical
$B$ and $V$ imagery from the KPNO 4-m prime-focus camera (Keel et al. 1999),
and deep $K$ and narrowband (redshifted H$\alpha$) imagery of the immediate
environs of 53W002 (Keel et al. 2002). 

An additional,
shallower $K$-band mosaic of the entire full-exposure ISO field was obtained during
one of the same NASA IRTF observing sessions. The NSFCam system was used 
at 0\farcs3 per pixel, for a typical exposure of 540 seconds for regions outside
the deep central observations. The region was covered with nine
$3 \times 3$ subrasters using 12" offsets. In many cases, the overlap
regions of these subrasters do not include bright enough objects to
register them from the $K$ data, so we stacked the images using 
relative astrometry from optical KPNO CCD images. The intensity scale
was set from observations of UKIRT faint standard stars (Hawarden et al. 2001).

A collective mean-flux limit was set to the submillimeter brightness of
the compact objects in the 53W002 grouping, by coadding the archival
SCUBA data, as discussed in section 3.3.

\section{Members of the 53W002 Group or Cluster}

\subsection{The radio galaxy 53W002}

The ISO data suggest a detection of the radio galaxy 53W002 itself at the
$\approx 3 \sigma$ level. Depending on the aperture size and details of
background subtraction, the derived flux is 0.024--0.039 mJy
or F$_\lambda$ = $(2.1 \pm 0.5) \times 10^{-19}$ 
erg cm$^{-2}$ s$^{-1}$ \AA$^{-1}$. We include a term for error due
to ``sky" fluctuations as evaluated in section 2.1, added in quadrature. This
is slightly conservative, in the sense that the error range 0.024--0.039 mJy
includes some of the effects of noise in the sky annulus. Thus, we adopt
a measurement of F$_\lambda$ = $(2.1 \pm 0.8) \times 10^{-19}$ 
erg cm$^{-2}$ s$^{-1}$ \AA$^{-1}$.
This furnishes a constraint on the continuum slope, and
therefore age, of the dominant stellar population.
The central AGN contributes a significant amount of light
at shorter wavelengths, both in emission lines and from its
continuum, which can be estimated from spectroscopy and
HST image analysis. Concentrating on the longer wavelengths
which are less affected by the most recent star-forming
history, we adopt starlight fractions of 0.77 at F110W, 0.62 at
F160W, and 0.78 in K (from PSF fitting in WFPC2 and NICMOS
images by Windhorst et al. 1998 and Keel et al. 2002, and
consistent with the spectroscopic estimate
of an optical spectral index $\alpha \approx 0.8$ for the AGN 
by Windhorst et al. 1991). This correction is small enough not
to add to the color error, dominated by the ISO point.
The starlight flux gives a ratio in F$_\lambda$ of $5.4 \pm 2.8$
between emitted wavelengths of 0.65 and 1.97$\mu$m. This translates to
emitted $R-K= 2.3 \pm 0.4$, or, adding the flat continuum shape from 
Keel et al. (2002),
$V-K= 3.2 \pm 0.4$.

The emission-line equivalent widths, and image decomposition of the
HST data, indicate that light from the active nucleus is a
significant contributor in the emitted-frame ultraviolet and
optical bands (Windhorst et al. 1991; Windhorst, Keel \& Pascarelle 1998). 
We thus examine whether either the AGN contribution
(perhaps reddened) might contribute to the observed
ISO flux. 
Line emission should contribute only a small fraction
of the ISO detection, based on scaling from the observed
Lyman $\alpha$, H$\alpha$, and [O III] emission.
At $z=2.4$, the response of the LW2 filter extends from 
1.47--2.50$\mu$m at half-peak transmission. This band contains
spectral features including Paschen $\alpha$ at 1.87$\mu$m,
as well as Br $\gamma$ at 2.16$\mu$m and the adjacent H$_2$
S(1) line, which is comparably strong in some galaxy contexts. Using the
constants in Osterbrock (1989, Table 4.4), for low densities at $10^4$ K,
the theoretical ratios to H$\alpha$ are 0.12 for Paschen $\alpha$ and
0.03 for Brackett $\gamma$. The IRTF data in Keel et al. (2002)
give a total intensity for H$\alpha$ (including some contribution from
[N II]) of $1.7 \times 10^{-15}$ erg cm$^{-2}$ s$^{-1}$. To convert the
prediction from scaling based on H$\alpha$ intensity to an in-band
LW2 flux, we take a simple trapezoidal approximation to the passband
shape, giving an approximate detected flux of 
$2.0 \times 10^{-15} \mbox{erg cm}^{-2} \mbox{s}^{-1}$
in the LW2 passband. Of this, the strong hydrogen
recombination lines would contribute at most about 15\%, so that
within these error bars we can safely neglect this correction.

There is a foreground elliptical galaxy at $z \approx 0.6$ appearing
about 7" NW of 53W002, potentially confusing the detection of
53W002. This is the galaxy numbered 6 by Windhorst et al. (1991, 1994), not
to be confused with object 6 of Pascarelle et al. (1996) and later
papers (which is 53W002 itself). 
We considered whether blending with this
object might increase the measured flux of 53W002 in the ISO data,
but its effect must be unimportant for two reasons.
The statistics of the data stack allow the possibility of two
sources at such a separation, but this would be inconsistent with
the astrometry of other sources nearby. In particular, the stellar
identification to the NW of 53W002 would become a blank field.
This is important because this object is so close that it would share
whatever detailed pointing errors apply to the area around 53W002 itself.
The star image is close to the expected PSF for a point source, which
implies that 53W002 (located 85" away) did not suffer pointing errors
between rasters which could degrade the image size. Furthermore,
the LW2 passband falls well on the long-wavelength side of the
stellar emission for even old populations at moderate
redshifts; we see this galaxy near $4.2 \mu$m in its emitted frame,
so the predicted flux is a small fraction of the observed value
for 53W002.

Fig. 4 shows the broadband photometry of 53W002 compared to
population synthesis models from Worthey (1994) for ages 1.0--2.0 Gyr, after 
allowance
made for the contribution of the central AGN following Windhorst et al. 
(1998) and Keel et al. (2002). Normalized to the near-IR points that
span the emitted optical range 4700--6600 \AA\ , these models indicate that
the error range on the ISO measurements spans the range 0.8--2.1 Gyr
in single-burst age, with the error range centered near 1.5 Gyr.
These values indicate a ``formation redshift" for the onset of
widespread star formation $z_f=3.6-7.7$, with the error range
centered at $z_f=4.7$. We neglect reddening and any contribution of
the AGN at 6.7$\mu$m deliberately, to get a maximum possible stellar
age. Even this measurement, of quite modest precision, narrows the
epoch of widespread star formation in 53W002 to the redshift range
oevr which galaxies are now detected.
Age estimates from mid-IR data should be
more robust than estimates from shorter-wavelengths, which
suffer from the bias toward the youngest populations as observed
in the emitted ultraviolet, and the potential role of dust
associated with the molecular gas (Scoville et al. 1997). Whether this 
advantage is realized in practice depends on the accuracy of the mid-IR data.
We note as well that there is lingering uncertainty over the near-IR role of
AGB stars in the 0.2-2 Gyr age range (as highlighted by Maraston 2004),
for which synthesis libraries are still poor in empirical tests.

Significant star formation in 53W002 took place over a span of at least 1 Gyr.
The emitted-optical spectrum from Motohara et al. (2001) shows a
substantial Balmer jump, indicating a stellar population
less than about $5 \times 10^8$ years old; a more complete,
self-consistent model would be usefully constrained only by
much smaller errors in the mid-IR data (as expected from forthcoming 
{\it Spitzer} results).
Thus we find that the
stars in 53W002 were formed over the cosmically brief but dynamically
significant span of 1--2 Gyr. In keeping with the general picture
of more massive objects beginning star formation first and being
able to bind their own enriched gas, 53W002 was the first luminous
system to begin star formation in this group, in contrast to the
lower-luminosity Lyman $\alpha$ emitters (section 3.3).

\subsection{Nondetection of a reddened AGN in a submillimeter-bright object}

Among the five known active nuclei in the 53W002 grouping, one
is also identified as a submillimeter-bright source.
Smail et al. (2003) have combined SCUBA mapping with VLA continuum
data at 1.4 Ghz to identify object 18 of Pascarelle et al.
(1996) with the submillimeter source, implying a bolometric
luminosity of $8 \times 10^{12}$ solar luminosities. The
origin of this energy output is not clear; all the properties of
this object observed in the optical and near-IR are traceable to
a luminous, largely obscured, active nucleus and its surrounding
emission and reflection nebulae. This interpretation is strengthened
by the {\it Chandra} spectrum, which shows a heavily absorbed
power-law form (White et al. 2004).
This object has an especially extensive
Lyman $\alpha$ halo, whose ionization is most easily explained if
our line of sight suffers unusually high extinction (so that the
UV continuum that we do observe would be scattered light). The ISO
nondetection of this object shows that any such luminous central
source must be quite strongly obscured. It must be fainter than
53W002 itself at 6.7$\mu$m, which requires an emitted-frame
color $V-K < 3.3$. Such a blue limit implies that any obscured
nucleus does not yet dominate the light at 2$\mu$m.
A typical QSO continuum must be reddened by $A_V > 1.2$ to have this 
color in $V-K$,
and the reddening must be substantially greater in this case
because the continuum at shorter wavelengths is quite blue (as shown
by Keel et al. 2002) and thus cannot be reddened AGN light (although scattered
light could play a role). The resolved continuum from NICMOS observations
also indicates that little direct AGN light emerges in our direction
in the $0.3-0.6 \mu$m emitted range. Given the uncertainties in
separating any direct starlight from side effects of the AGN, we
can limit direct light from the AGN at these shorter wavelengths to
less than about 20\% of the total. 
 
\subsection{Narrow-Line Lyman $\alpha$ Star-Forming Objects}

We set a mean limit on the $6.7 \mu$m flux of the faint blue
Lyman $\alpha$ emitters in this field, by adding $20 \times 20$" regions
around each of them registered on the optical positions. The objects
stacked were numbers 5,11,12,29,34,60, and 113 from
Pascarelle et al. (1996). Of these, 5 and 12 are at $z=2.05$, 94 is at
$z=2.74$; and the rest are at $=2.30-2.40$ (Pascarelle et al. 2004).
This
mean image gives a flux formally less than 0.010 mJy,
slightly better than the noise statistics should allow.
The 1$\sigma$ noise level for 10" apertures averaged over 7 objects gives 0.003
mJy,
and the 5"-diameter effective apertures used for these unresolved objects would
give group-mean
1$\sigma$ levels close to 0.004 mJy. We therefore take the more
conservative $3 \sigma = 0.012$ mJy value, since the measured formal
flux might be unusually low by small-number statistics. This means 
that on average they are at least three times fainter than 53W002 itself.
The mean spectral slope can be evaluated against observed wavelengths
near 1.6$\mu$m using NICMOS data (Keel et al. 2002),
using the sample mean flux of $1.4 \times 10^{-19}$ erg cm$^{-2}$ s$^{-1}$
\AA$^{-1}$. The mean $F_\lambda$ flux ratio between 
emitted wavelengths 0.47 and 2.0$\mu$m is then $>2.5$, implying $V-K < 4.0$.
This means that these objects could in principle have stellar
populations as old, and formed over a span as long,
as we infer for 53W002. A limit to dust emission from ongoing
star formation can be set from the coadded SCUBA data, likewise stacked
at the positions of these objects. The formal mean 850 $\mu$m flux
is $0.14 \pm 0.24$ mJy. For objects with a spectral energy distribution like
that of Arp 220, this flux (0.14 mJy) at $z=2.4$ would correspond to a total
far-IR luminosity $\approx 3 \times 10^{11}$ solar luminosities, or
a mean star-formation rate of about 30 solar masses per year. This
is a few times lower than implied by the typical mid-UV continua of
these objects (using the star-formation form from equation 1 of
Kennicutt 1998), which may attest to the nearly dust-free 
nature of the star-forming regions as seen in the very blue
UV-optical colors (Keel et al. 2002).


In themselves, then, these data do not
add to the case that the compact Lyman $\alpha$ emitters
 arise from short (perhaps repeated)
bursts, either in small ``protogalactic" systems
(Pascarelle et al. 1996, 1998; Keel et al. 2002) or
as pieces of larger systems whose typical surface brightness is
below current detection thresholds (Colley et al. 1996).


\section{Additional galaxy detections}

Two lower-redshift galaxies are certain or probable detections at
$6.7\mu$m. The brightest, strong enough to be important in the
astrometric registration of the data, is the merging system
53W003 at $z \sim 0.05$ (from the
LBDS followup spectrum described by Windhorst, Kron, \& Koo 1984). 
This system is a modest 
radio source as
well (Windhorst, van Heerde, \& Katgert 1984, from their table of additional
sources). The detailed structure at 
$6.7 \mu$m is reproduced between two overlapping
individual rasters, separating the northern and southern nuclei
and showing the southern nucleus about twice as bright as the
northern one. Convolved to similar resolution, both nuclei
have similar $B-V$ and a flux ratio of about 1.4 (south/north).
This object is identified with the IRAS faint-source survey
object IRAS F17130+5021, with catalogued fluxes 
$F_{12} < 0.07$ Jy, $F_{25} < 0.08$ Jy, $F_{60} = 0.31$ Jy,
and $F_{100} = 0.86$ Jy. This spectral shape is not particularly
``hot" for a merging system, so it is either relatively
quiescent in star formation or has not yet initiated
a major starburst.

One fainter source, in the inner region of full exposure, has a 
likely counterpart in a spiral galaxy
at $z=0.3-0.6$, which appears in the
WFPC2 images from Pascarelle et al. (1996).  This spiral has two 
small, close, and blue
companions. 
This is the optically brightest disk galaxy within the
region of full ISO exposure. HST and ground-based images of these
galaxies from $B$-$K$ are shown in Fig. 5.

The optical and near-infrared magnitudes for these galaxies are
listed in Table 2. We have estimated a redshift range for the faint
spiral from its colors,
using the $K$-correction calculations by Kinney et al. (1996).

One additional ISO source, slightly brighter than 53W002, falls near
a faint galaxy (source number 7 in Fig. 3 and Table 1). This 
galaxy lies outside the WFPC2 and NICMOS observations of this
region, so we have only ground-based broadband color data.
Within a 3" aperture it has $B=23.16$, $V=21.75$, making it
slightly brighter and significantly redder than 53W002 itself (which has
$V=22.6$ and $B-V \approx 0.8$).

\section{Mid-Infrared Source Counts}

Source counts are of interest from any deep observation in
newly-opened wavelength bands. The LW2 band around 6.7$\mu$m falls in an
interesting spectral range. At low redshifts, this is
on the long-wavelength, nearly Rayleigh-Jeans tail of emission from
most stellar photospheres, so that only quite bright local galaxies
(plus cool Galactic stars) will be prominent. At large redshifts,
photospheric radiation shifts into this passband, making optically
much fainter objects detectable. There will be a ``valley" at moderate 
redshifts (from a few tenths to about $z=2$) in which the only strong 
sources will be AGN, with flatter spectra sometimes augmented by hot dust.
Since only a single ISO source appears to be associated with the 
targeted group at $z=2.4$, we can use the others as a fair sample of the
deep extragalactic sky. For the small solid angle covered, these
counts serve largely as a sanity check on our reduction.

This region is part of the Herc 2 field surveyed in several passbands during
the Leiden-Berkeley Deep Survey (LBDS; Windhorst, van Heerde, \& Katgert 1984;
Windhorst, Kron, \& Koo 1984; Kron, Koo, \& Windhorst 1985). 
The extinction is modest
given the rather low galactic latitude $b=35^\circ$, with estimates
from $A_B=0.03$ from Burstein \& Heiles (1984) to 0.09 (Schlegel et al.
1998). The far-IR cirrus emission and H I column density in this 
direction are similarly moderate for its latitude. Optical and radio
source counts in this region reach very deep, with substantial 
numbers of photometric measurements and redshifts available.

Several additional deep studies were carried out using ISOCAM at 6.7$\mu$m, 
facilitating a comparison of source counts and populations.
We compare here results from the HDF observations of 
Oliver et al. (1997) and Serjeant et al. (1997), the HDF-S observations 
analyzed by Oliver et al. (2002), the ELAIS data presented by Serjeant et al. 
(2000), the SSA13 deep survey analyzed by
Taniguchi et al. (2002) and Sato et al. (2003), the
lens-amplified counts behind Abell 2390 from Altieri et al. (1999),
and the Lockman Hole data from Taniguchi et al. (1997).
Our coverage in solid angle is about 25500 arcsec$^2$ or
$5.86 \times 10^{-7}$ sr at full exposure, with the envelope
of the entire area with any exposure spanning 89283 arcsec$^2$ or
$2.05 \times 10^{-6}$ sr. The strongest source, 53W003 at 7.0
mJy within a 15" radius, is brighter than would be expected in
a random field based on the ELAIS+HDF counts at this wavelength
from Serjeant et al. (2000), at about the 95\% level. Counts of the fainter 
sources, limited as they are in number, are in reasonable agreement with
the other surveys (Fig. 6). 

We consider three subsamples of our source list. One subsample (labelled
``ALL") 
consists of all
except 53W002, which was the targeted object and should be omitted
in constructing a random sample. The minimum list (``MIN") includes sources
with optical identifications, which should be free of spurious
IR souces at the expense of rejecting genuine sources which are
very faint at shorter wavelengths. A further refined minimal 
extragalactic list (``XGAL") includes only those sources with nonstellar 
optical counterparts. Fig. 6 compares the highest and lowest of these 
with published counts, in the cumulative form, showing good agreement. 
In particular, the
XGAL sample overlaps, within its error bounds, the ranges seen in
the HDF, Abell 2390, and the Lockman Hole data, while the SSA 13
counts fall somewhat below all of these. This may fit with the
radio-source counts; both the HDF-N and SSA13 are low compared to
another $\approx 12$ deep survey regions, in both cases connected
to explicit field selection against strong radio sources (see Windhorst 2003).
In retrospect, these higher counts are consistent with {\it Spitzer}
counts transformed from the 5.8$\mu$m results reported by Fazio et al.
(2004).

\section{Conclusions}

We have carried out a deep survey of the 53W002 galaxy group at
$z=2.4$, using ISO at roughly the emitted $K$ band, to detect
or limit any older stellar populations in the members and thus
probe the onset of significant star formation. The radio galaxy
53W002 was detected at the 3$\sigma$ level, giving emitted
$V-K$ colors consistent with the earliest widespread star
formation setting in at $z=3.6-7.7$. The neighboring 
star-forming objects detected as narrow Lyman $\alpha$ emitters are
so faint that even a group average flux limits provides a less
stringent constraint than this; their emitted UV and implied
composition give stronger constraints on their star-forming
history at this point. The neighboring submillimeter-bright
AGN is undetected in the ISO data, indicating that the
direct continuum radiation from its active
nucleus is quite heavily absorbed.

All these issues will be substantially clearer with the analysis
of recent {\it Spitzer}
observations by the IRAC team. These data should be
able to narrow the onset of star formation in all these objects,
and hence tell whether the more massive objects indeed began
star formation earlier. These data might be able to test,
at early times, the role of such issues as the upper asymptotic-giant
branch in populations 0.2--1 Gyr old, whose potential importance
in the near-IR bands has recently been stressed by Maraston (2004).

\acknowledgments
This work was supported by NASA through JPL contract 961525.
We thank Dave van Buren for suggestions on observing strategies for
achieving the best sensitivity. The ISOCAM data presented in this
paper were analyzed using ``CIA", a joint development by the ESA
Astrophysics Division and the ISOCAM Consortium.
Stephan Ott and Babar Ali provided crucial assistance in
setting up the ISOCAM analysis and in dealing with special
problems that arose in processing these observations.

\clearpage

\vfill\eject

\begin{table*}
\begin{center}
\begin{tabular}{rcccl}
\tableline
\tableline
ID & $\alpha_{2000}$ & $\delta_{2000}$ & F$_\nu$, mJy & Notes \\
\tableline

 1 & 17:14:16.9 & +50:18:16 & $6.5 \pm 1.1$     & 53W003\\
 2 & 17:14:04.8 & +50:17:22 & $0.080 \pm 0.010$ & 5" W of star\\
 3 & 17:14:19.1 & +50:17:20 & $0.069 \pm 0.006$ & star\\
 4 & 17:14:05.3 & +50:15:17 & $0.069 \pm 0.015$ & blank field\\
 5 & 17:14:21.5 & +50:16:17 & $0.057 \pm 0.008$ & blank field\\
 6 & 17:14:10.2 & +50:16:08 & $0.045 \pm 0.006$ & red star\\
 7 & 17:14:17.2 & +50:15:04 & $0.044 \pm 0.006$ & faint galaxy?\\ 
 8 & 17:14:15.2 & +50:16:50 & $0.041 \pm 0.005$ & red star or compact galaxy\\
 9 & 17:14:12.8 & +50:16:45 & $0.035 \pm 0.006$ & spiral galaxy\\
10 & 17:14:14.8 & +50:15:30 & $0.031 \pm 0.007$ & 53W002\\

\tableline
\end{tabular}

\tablenum{1}
\caption{
Source detections in LW2 field\label{tbl1}}
\end{center}

\end{table*}

\begin{table*}
\begin{center}
\begin{tabular}{rccccccl}
\tableline
\tableline
Source & $B$ & $V$ & $I$ & $K$ & aperture & $z_{phot}$ & Notes \\
\tableline
 1 & 15.85 & 14.78 &   ... & ... & 32"  & 0.05     & 53W003 \\
 9 & 19.17 & 17.56 & 16.61 & ... & 4\farcs0 & 0.3--0.6 & \\
\tableline
\end{tabular}

\tablenum{2}
\caption{
Optical fluxes and redshifts for detected galaxies \label{tbl2}}
\end{center}
\end{table*}

\clearpage

\figcaption
{Overlap pattern of individual LW2 rasters superimposed over
a version of processed 6.7$\mu$m mosaic, in which no masking of
raster borders was done so as to preserve the full field for this
illustration and background variations have not been removed. The numbers
denote observation sequence numbers; 931--947 are from orbit
500, while 703--729 were observed on orbit 521. The circled cross
shows the location of 53W002. The low-redshift system 53W003
is prominent to the northeast (upper left).
\label{fig1}}

\figcaption
{Final mosaicked 6.7$\mu$m image of the 53W002 field, incorporating 
masking of the edges of individual rasters to reduce artifacts in
the clipped mean and median filtering of sky variations though
an $11 \times 11$" window. The rapid increase in noise to the edge of the
mosaic, outside the central 90" radius, is still apparent. The maximum
extent of the exposed region is 347" N-S and 354" E-W. The prominent
elongated source at upper left is the low-redshift merging system
53W003. 
\label{fig2}}

\figcaption
{Sources in the 53W002 field, as shown on a Gaussian-smoothed
LW2 image (smoothed to 5" FWHM, left) and a blue-light 
image (right, with 4100-\AA  data from Keel et al. 1999) resampled
to the same scale.
Sources listed in Table 1 are marked. Number
1 is 53W003, number 10 is at the position of 53W002 at
$z=2.39$.
\label{fig3}}

\figcaption
{The infrared spectral energy distribution of 53W002 compared to model
stellar populations. The near-IR points, mapping to the
emitted optical near the $B$ and $R$ bands, are the NICMOS
and IRTF data from Keel et al. (2002)
incorporating correction for the central AGN. The model stellar
populations are from the Worthey (1994) code as implemented on the
World-Wide Web. The error range for the ISO point spans ages of
1.0--2.0 Gyr, corresponding to redshifts at the onset of significant
star formation $z_f=3.6-7.0$ with a central value of $z_f=4.7$. 
\label{fig4}}

\figcaption
{Optical images of moderate-redshift galaxies with ISO 6.7$\mu$m
detections. The images of 53W003 are B and V KPNO 4m prime-focus data
from Keel et al. (1999), since this systems falls outside our 
HST imagery. The southern nucleus of this apparently merging system
is the peak of 6.7$\mu$m radiation. The area shown for 53W003 is
67 arcseconds square.
For the spiral listed as ISO source 9,
$10 \times 10$-arcsecond sections of
WFPC2 images are shown in F450W (B), F606W (V), and
F814W (I), from the observations described by Pascarelle et al. (1996).
The $B$ image is reconstructed from data taken using a
$2 \times 2$ dither pattern so the pixel spacing is 0\farcs05 rather than
0\farcs1 as obtained from a single WFPC2 image (as for the $V$ and $I$ data).
The brightness scale in each case is pseudologarithmic, using a flux offset of
1\% of the maximum to avoid a discontinuity at the mean sky level.
Each image has been rotated to place north at the top and east to the left.
\label{fig5}}

\figcaption
{Counts of ISO sources from deep observations at 6.7$\mu$m.
Cumulative counts from our observations in the 53W002 field are shown for
all sources and for those objects not optically identified with
foreground stars (XGAL). The error bars are for Poisson statistics,
except that the highest-flux point, above which there is a single
detection, is shown with $\pm50$\% error bounds for clarity.
The shaded region encompasses the $\pm 1 \sigma$ bounds for the
XGAL and all-detection sample. These overlap substantially
with the counts reported for the HDF , Abell 2390, and Lockman Hole
data, especially for the XGAL sample.
\label{fig6}}

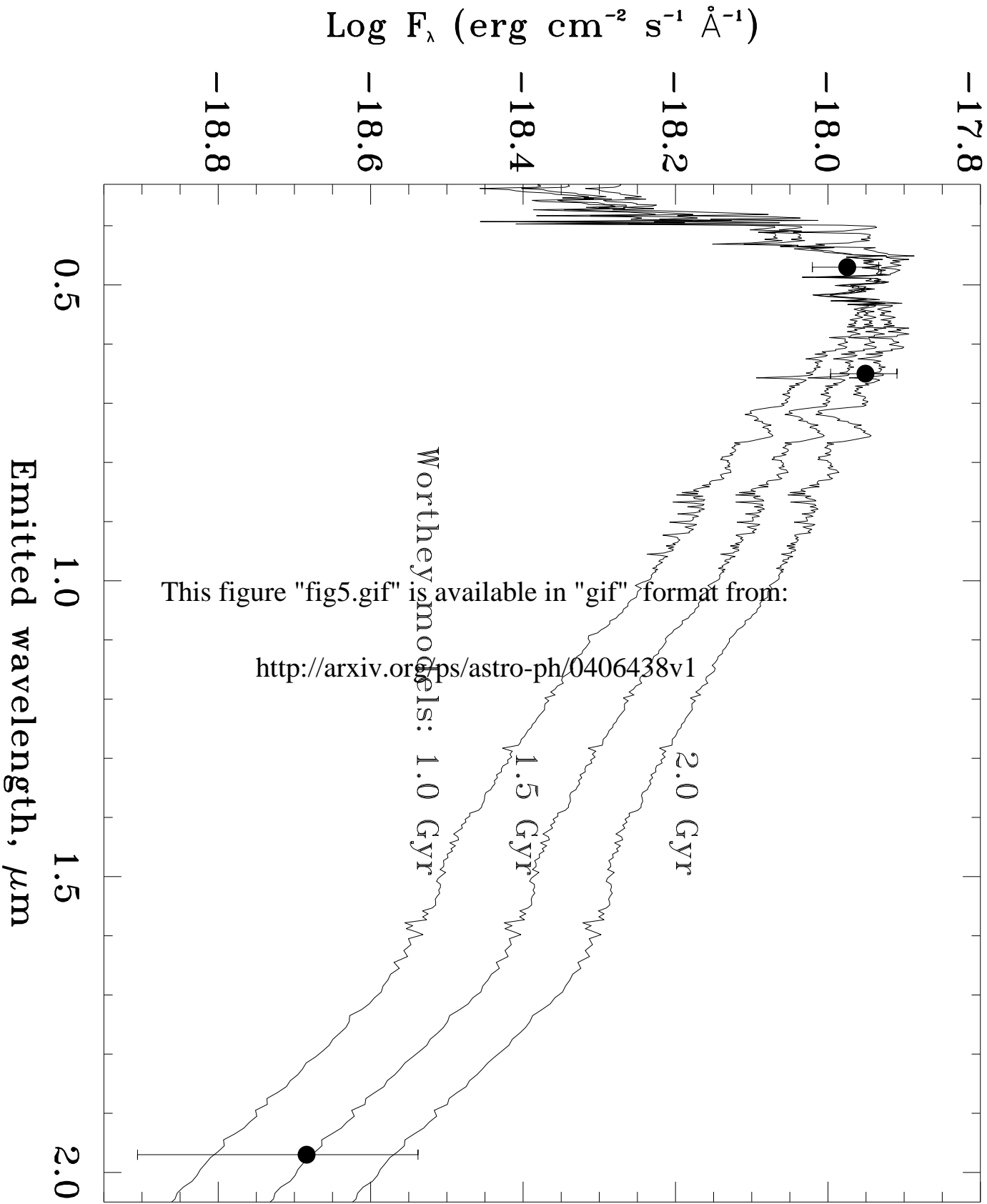
\end{document}